\documentclass{article}
\usepackage{arxiv}
\usepackage[utf8]{inputenc} 
\usepackage[T1]{fontenc}    
\usepackage{hyperref}       
\usepackage{url}            
\usepackage{booktabs}       
\usepackage{amsfonts}       
\usepackage{nicefrac}       
\usepackage{microtype}      
\usepackage{lipsum}
\usepackage{xcolor}
\usepackage{amsmath,amstext,amssymb}
\usepackage{graphicx,verbatim,float,subfigure,graphicx}
\graphicspath{ {./images/} }

\usepackage{amssymb}
\usepackage{svg}
\usepackage{multirow}  
\usepackage[export]{adjustbox}
\usepackage{soul}
\usepackage{textcomp}
\usepackage{multirow}
\usepackage{enumitem}

\title{Microscale Morphology driven Thermal Transport in Fiber Reinforced Polymer Composites}

\author{
  Sabarinathan P Subramaniyan \\
  Dept. of Mechanical Engineering \\
  University of Wisconsin-Madison \\
  Madison, WI 53706 \\
  \And
  Jonathan D Boehnlein \\
  Dept. of Mechanical Engineering\\
 University of Wisconsin-Madison \\
  Madison, WI 53706 \\
  \And
 Pavana Prabhakar \\
 Dept. of Mechanical Engineering \\
  Dept. of Civil \& Env. Engineering \\
  University of Wisconsin-Madison \\
  Madison, WI 53706 \\
  \texttt{pavana.prabhakar@wisc.edu} \\
}

\date{ }

\begin{document}
\maketitle
\begin{abstract}
Fiber-reinforced polymer composite (FRPC) materials are used extensively in various industries, such as aerospace, automobiles, and electronics packaging, due to their remarkable specific strength and desirable properties, such as enhanced durability and corrosion resistance. The evolution of thermal properties in FRPCs is crucial for advancing thermal management systems, optimizing material performance, and enhancing energy efficiency across these diverse sectors. Despite significant research efforts to develop new materials with improved thermal properties and reduced thermal degradation, there is a lack of understanding of the thermal transport phenomena considering the influence of microscale reinforcement morphology in these composites. In the current study, we performed experimental investigations complemented by computations to determine the thermal transport properties and associated phenomena in epoxy and carbon fiber-reinforced epoxy composites. The experimental findings were utilized as input data for numerical analysis to examine the impact of fiber morphology and volume fraction in thermal transport phenomena. Our results revealed that composites incorporating non-circular fibers manifested higher thermal conductivity than traditional circular fibers in the transverse direction. This can be attributed to increased interconnected heat flow pathways facilitated by the increased surface area of non-circular fibers with the same cross-sectional areas, resulting in efficient heat transfer.

\end{abstract}



\section{Introduction}\label{intro}
Fiber-reinforced polymer composite (FRPC) materials have become increasingly widespread in various industries, ranging from aerospace and automotive to civil and marine. This can be attributed to their notable specific strength, improved durability, and exceptional corrosion resistance. Moreover, FRPC materials have also been utilized in the electronics sector, specifically in thermal management. By manipulating key parameters such as reinforcement type, reinforcement volume fraction, and polymer network, the thermal properties of polymer composites can be customized to meet insulative and high heat dissipative requirements. In this paper, fiber cross-section shape is considered as a parameter to modulate the thermal properties of FRPCs particularly focusing on non-circular fiber cross-sections.

There have been few prior works on understanding the role of non-circular fiber cross-sections on the mechanical properties of FRPCs. Herraz et al.\cite{herraez2016computational} conducted a numerical investigation to elucidate the impact of the cross-sectional morphology of fibers on the transverse mechanical response of unidirectional fiber-reinforced composite materials. Their simulation outcomes revealed a 10\% increase in tensile strength for fibers characterized by 2-lobed and elliptical cross sections when subjected to tension along the fiber alignment axis. Conversely, a reduction ranging between 12-25\% was observed in the tensile strength perpendicular to the fiber alignment direction. Furthermore, the lobular configuration of fiber cross sections demonstrated an increase of 20\% in compressive strength relative to the conventional circular fiber morphology. This investigation identified two principal mechanisms governing the observed damage phenomena: interfacial debonding and the formation of shear bands within the matrix material.
Similarly, Kitagawa et al.\cite{kitagawa2022experimental} undertook a comprehensive examination into the impact of fiber cross-sectional morphology on the transverse crack propagation behavior of carbon fiber reinforced polymer composites, employing a combined experimental and computational approach. Their investigation elucidated the underlying mechanisms governing transverse crack extension. Specifically, they observed that conventional circular-shaped carbon fibers tend to experience increased crack propagation facilitated by interfacial debonding, whereas in the case of kidney bean-shaped fibers, matrix damage predominantly drives crack extension. Consequently, the researchers inferred that the deployment of kidney bean-shaped carbon fibers delays the onset of transverse cracking and mitigates their cumulative occurrence within the composite material.
Oak Ridge National Laboratories (ORNL) researchers have studied two types of low-cost carbon fibers, namely kaltex (kidney bean-shaped fiber cross-section) and Taekwang fiber (circular fiber cross-section). This innovation aims to facilitate their integration into wind turbine design, particularly for the construction of larger rotors, thereby enhancing energy efficiency within the domain of wind power generation. Later, a comprehensive analysis was conducted to assess the mechanical performance of low-cost carbon fiber composites, with Industrial carbon fiber Zoltek PX35 serving as the benchmark material. This study yielded significant findings, particularly regarding the fatigue resistance of the low-cost composites. In stark contrast to their commercial counterparts, which experienced a substantial reduction in maximum tensile strain after numerous fatigue cycles, the low-cost composites demonstrated remarkable durability, maintaining consistent levels of maximum tensile strain even after extensive fatigue testing. Notably, while the tensile strength of the low-cost composites exhibited a reduction of approximately 43\% compared to the commercial variants, the compressive strength and modulus remained largely unchanged. \cite{ennis2019optimized,miller2019mechanical}

With regard to thermal properties, previous studies have primarily concentrated on increasing the thermal conductivity of polymer composites by incorporating various reinforcement types, such as graphite, boron nitride, and carbon nanotubes.\cite{Mehra2018, Guo2020, han2020enhanced, huang2020bi}. The influence of fiber cross-sectional morphology on mechanical properties was investigated by Ye et al.\cite{yeinfluence}. Their study focused on different morphologies of glass fibers, including circular, peanut-shaped, and oval-shaped fibers. It was observed that fibers with larger aspect ratios exhibited a reduction in transverse moduli. Xu et al.\cite{xu2008effect} studied the surface characteristics of kidney bean-type and circular cross-sections in the composite interface. Their investigation revealed that kidney bean-shaped fiber/epoxy composites exhibited superior performance in terms of interfacial shear and interlaminar shear loads compared to circular fiber/epoxy composites, with improvements of 23.5\% and 12.7\%, respectively. A previous fundamental study performed by the current authors \cite{subramaniyan2021fiber} showed that moisture diffusion is significantly impacted by fiber morphology and volume fraction in fiber-reinforced composites. Although the thermal and moisture transport properties have significant implications, previous research has predominantly concentrated on investigating the mechanical performance aspects. 

This paper examines the interplay between fiber cross-sectional morphology, volume fraction, and thermal transport distribution within polymer composites, focusing on thermal transport properties. Despite research efforts to develop new materials to improve thermal properties, a comprehensive understanding of thermal transport phenomena still needs to be developed, considering the influence of microscale reinforcement architecture in fibrous composites. To that end, experimental investigations complemented by computations to determine the thermal transport properties and associated phenomena in epoxy and carbon fiber-reinforced polymer (CFRP) composites were performed. Thermal diffusivity, thermal conductivity, and specific heat capacity were measured experimentally for epoxy and CFRP composites with circular fibers fabricated in this work. These experimental findings were then used as input data in the computational modeling to examine the impact of fiber morphology (non-circular) and volume fraction on thermal transport phenomena.



\section{Experimental methods}

\subsection{Materials and Manufacturing}

In this study, carbon fiber-reinforced epoxy composites are fabricated through the wet layup method. Specifically, PYROFIL TR50S carbon fibers and IN2 Epoxy Infusion Resin with AT30 slow hardener obtained from easy composites Ltd are used. The ratio of epoxy resin to hardener is kept at 100:30 and cured for 24 hours at room temperature. Table~\ref{tab:mat_props} shows the typical properties of each constituent material used in this paper.

\begin{table}[H]
\centering
\renewcommand{\arraystretch}{1.2}
\begin{tabular}{ccc}
\hline
\textbf{Property}        & \textbf{Carbon fiber} & \textbf{Epoxy} \\ \hline
Density                  & 1.82 g/cm3            & 1.11 g/cm3     \\ 
Thermal   conductivity   & 7 W/m.K*              & 0.124 W/m.K**  \\ 
Specific heat   capacity & 740 J/Kg/K            & 1160 J/Kg/K**  \\ \hline
\end{tabular}
\caption{Material Properties of individual constituents that are employed in the numerical modeling\cite{baril2013improving}}
\vspace{0.01in}
** experimentally determined \\
* manufacturer specified
\label{tab:mat_props}
\end{table}

\subsection{Volume Fraction - Thermogravimetric analysis}
The fiber volume fraction of the fiber-reinforced polymer composites mentioned in the previous section is determined using the widely employed Thermogravimetric analysis (TGA). Determining fiber volume fraction in composite materials is critical for their characterization and performance evaluation. Details of TGA are provided in the Supplementary Document.

\subsection{Thermal transport characterization}
Thermal properties, like thermal diffusivity, thermal conductivity, and specific heat capacity, are measured using NETZSCH LFA 447 (a Laser flash analysis technique) according to ASTM E1461\cite{ASTM_laser}. Details of LFA are provided in the Supplementary Document. In this work, these properties are measured for epoxy samples and for CFRP composite samples in two directions, one along the fiber direction (longitudinal) and the other perpendicular to the fiber direction (transverse).

The experimentally measured fiber volume fraction of the manufactured CFRP composite and the thermal properties of epoxy and CFRP composite are used as input to the computational model described next.

\section{Computational methods - Thermal transport modeling}

The computational modeling focuses on determining the thermal conductivity of circular, bean-shaped, triangular, and bi-lobular fibers with varying fiber volume fractions. Here, the area of the fiber cross-sections is held constant at 38.48 $\mu m^2$ for all shapes. However, the perimeter of these shapes vary as 21.99 $\mu m$ for circular, 23.9 $\mu m$ for triangular, 24.04 $\mu m$ for bean, and 25.28 $\mu m$ for bi-lobular.

\begin{figure}[h!]
\centering
\subfigure[]{
\includegraphics[width=3.5cm]{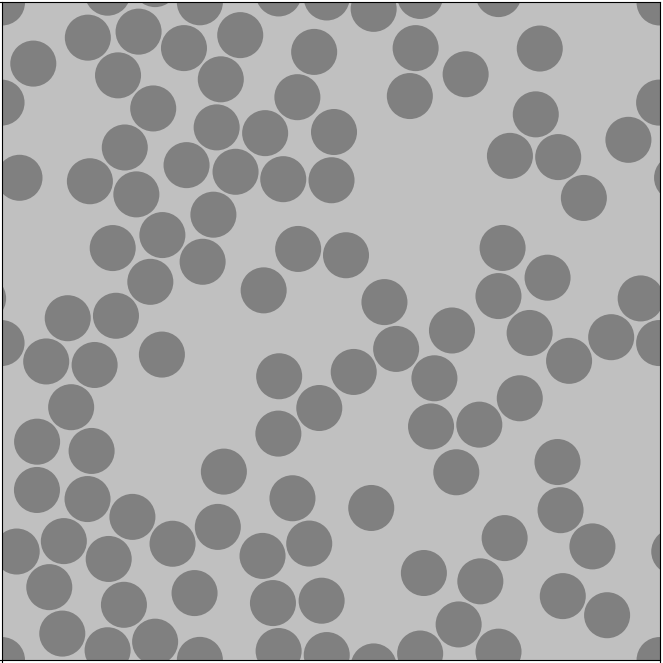}
}
\hspace{0.1in}
\centering
\subfigure[]{
\includegraphics[width=3.5cm]{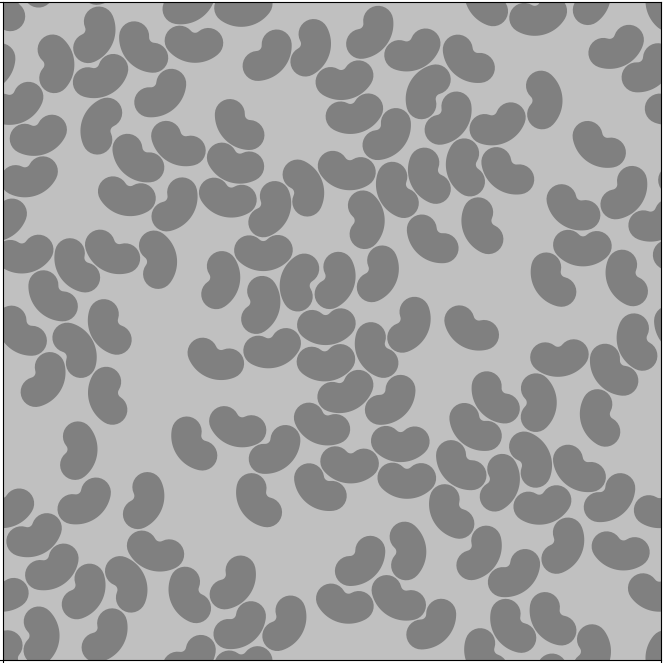} 
}
\hspace{0.1in}
\centering
\subfigure[]{
\includegraphics[width=3.5cm]{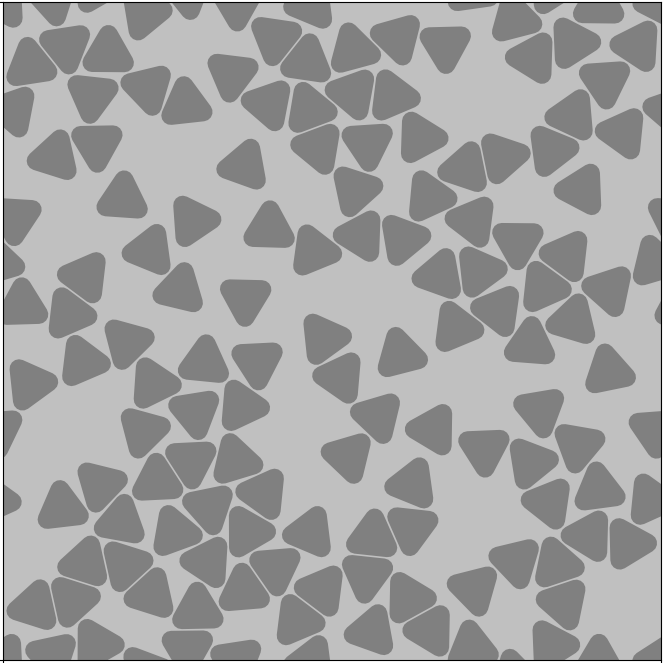} 
}
\hspace{0.1in}
\centering
\subfigure[]{
\includegraphics[width=3.5cm]{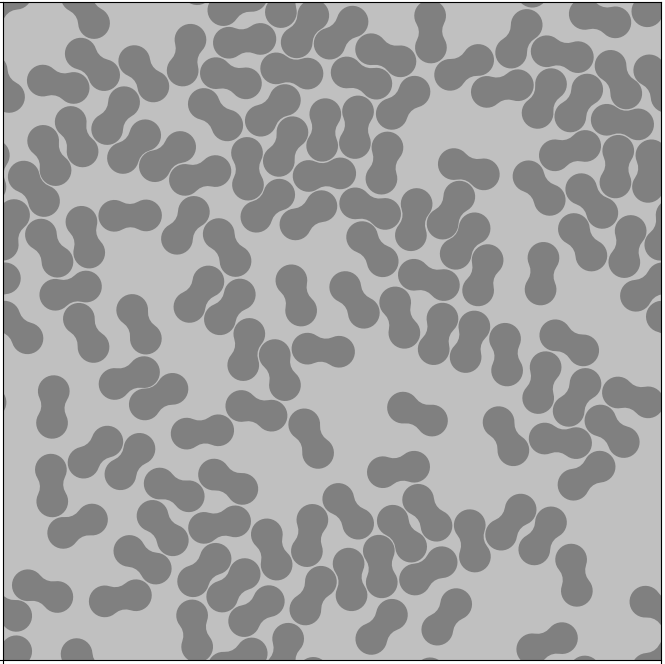} 
}
\caption{RVE to determine the coupled moisture and thermal transport properties with different fiber morphology(a) Circular, (b) Bean shaped, (c) Triangular (d) Bi-lobular}\label{RVE_shapes}
\end{figure}

\begin{figure}[h!]
\centering
\includegraphics[width=16cm]{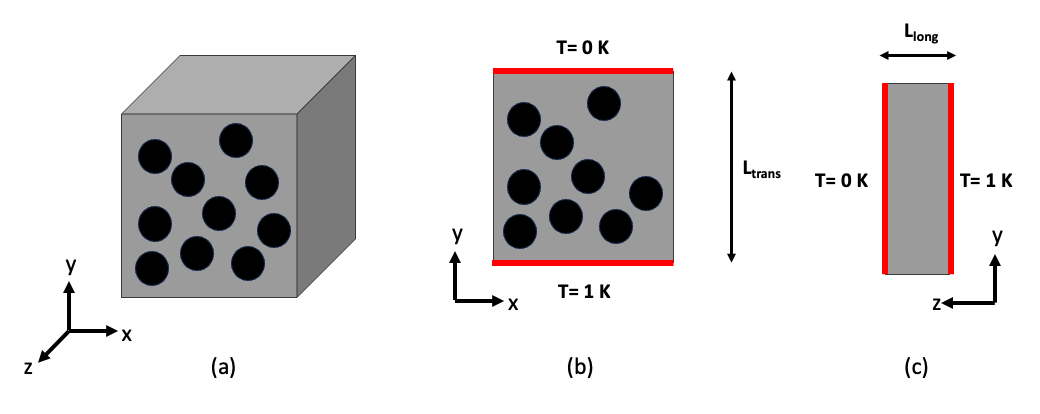}
\caption{Computational model domain for heat transfer analysis with prescribed boundary conditions}\label{heatTransferAbaqus}
\end{figure}

Fig~\ref{RVE_shapes} shows the RVE generated using the VIPER tool\cite{Herraez2020} for different randomly distributed fiber-shaped reinforcements. To solve the steady state heat conduction problem, the Heat Transfer module within the finite element analysis software, ABAQUS is utilized. Boundary conditions of T = 1 K and T = 0 K in the measurement direction (Figure~\ref{heatTransferAbaqus}(b) - perpendicular to fiber, Figure~\ref{heatTransferAbaqus}(c) - fiber direction) and no flux boundary conditions on all the remaining surfaces are prescribed. Material properties used in these models are listed in Table~\ref{tab:mat_props}.

The dimensions of the domain in the plane perpendicular to the fiber direction (X-Y) are set at 100 x 100 µm, which was determined as the optimal size based on our prior research \cite{subramaniyan2021fiber}.
Additionally, a parametric study is conducted to establish the appropriate thickness along the z-direction of the domain, resulting in a value of 1 µm.  Then, the effective thermal conductivity is calculated using $K_i = \frac{Q_i * L }{\Delta T} $, where $K_i$ and $Q_i$ represent the thermal conductivity and Heat flux along directions 1,2,3, respectively. $L$ represents the length along the heat flow direction and $\Delta T$ represents the temperature change.

\section{Results and Discussion}\label{res}

\subsection{Determination of fiber volume fraction - Thermogravimetric analysis}

To determine the fiber volume fraction in the manufactured CFRP composite, TGA was performed individually on pure epoxy and the composite, which resulted in the production of char residue. The neat epoxy had a residual char weight ($W_r$) content of approximately 3\% exhibited by a plateau region between 500 to 800 degrees Celsius. In a similar temperature range, the CFRP composite also produces a char residue ($W_c$=32\%), which is a combination of char from the epoxy matrix and the fibers, the latter being unaffected by heating in nitrogen environment. The weight fraction of fiber ($W_f$) is determined to be 30\%, whcih was calculated using Equation~\ref{Wf} from the supplementary document. From Equation~\ref{Vf} shown in the supplementary document, the fiber volume fraction is determined to be in the range of 20\% - 22\% for the CFRP samples fabricated in this work.

\subsection{Effect of temperature on thermal transport properties of epoxy and Carbon fiber reinforced epoxy composite - Experimental observations}

Heat transfer in insulating materials such as polymers is facilitated by generating discrete units of energy known as phonons, which are produced by atomic lattice vibrations. The thermal conductivity of polymer and polymer composite materials is directly related to the average distance that phonons travel before experiencing scattering, referred to as the mean free path. The mean free path plays an essential role in determining the thermal conductivity of these materials. The phonon mean free path is influenced by three distinct scattering types: phonon-phonon scattering, phonon-boundary scattering, and phonon-impurity scattering. Temperature is another factor that affects the phonon mean free path. As temperature increases, the dominant phonon wavelength decreases, increasing phonon-phonon scattering and reducing the phonon's mean free path, ultimately reducing the thermal conductivity. \cite{Mehra2018,Guo2020}

\begin{figure}[h!]
\centering
\subfigure[]{
\includegraphics[width=7.cm]{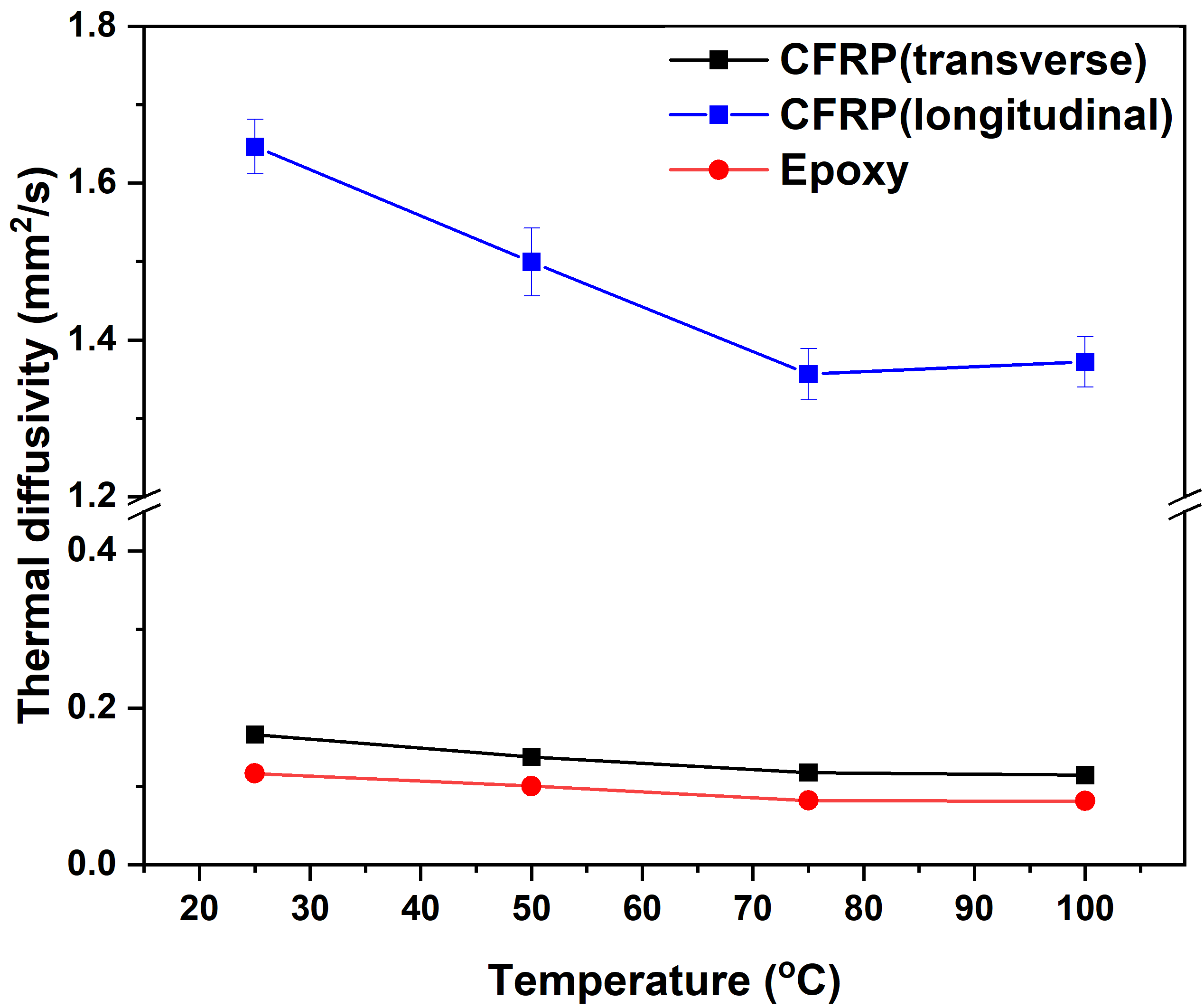}\label{diffusivity_unaged}
}
\hspace{0.2in}
\centering
\subfigure[]{
\includegraphics[width=7.cm]{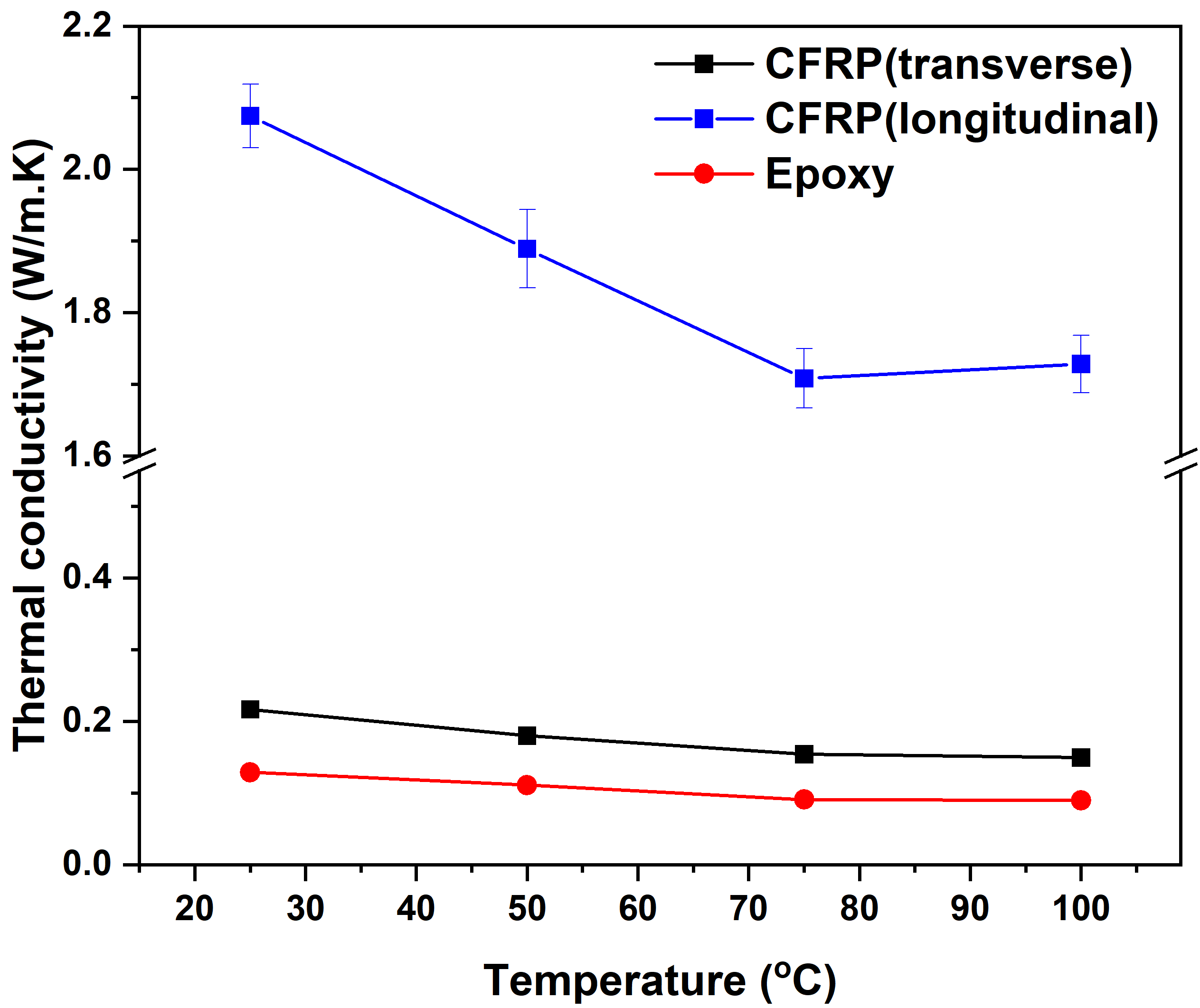}\label{conductivity_unaged}
}
\hspace{0.2in}
\centering
\subfigure[]{
\includegraphics[width=7.cm]{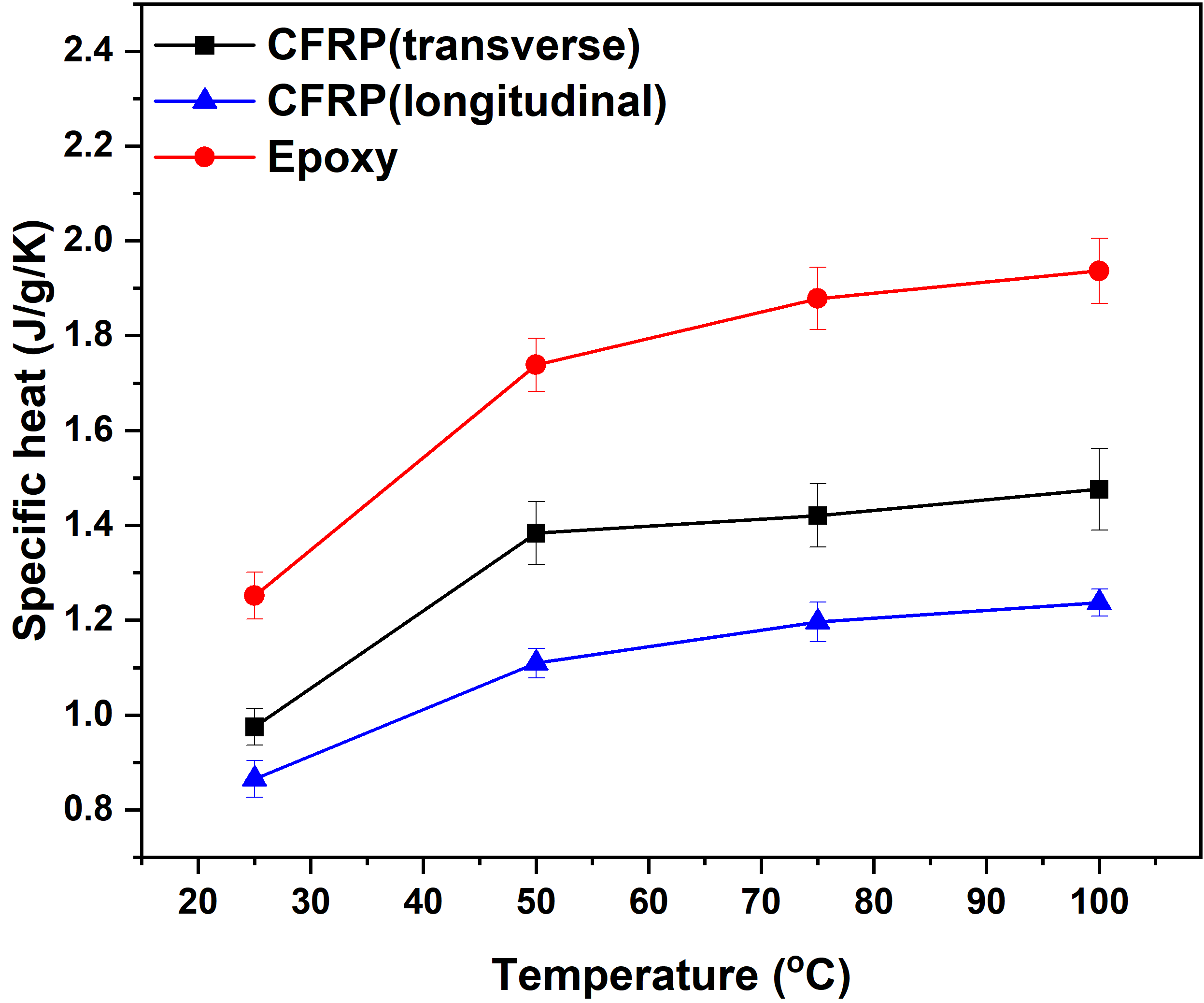}\label{specific_heat_unaged}
}
\caption{Influence of Temperature on thermal transport properties of epoxy and CFRP composites  (a) Thermal diffusivity (b) Thermal conductivity and (c) Specific heat capacity}
\end{figure}

Fig~\ref{diffusivity_unaged} and \ref{conductivity_unaged} show experimentally determined thermal diffusivity and thermal conductivity of the epoxy and CFRP composite samples manufactured in this work, as a function of temperature. In general, the incorporation of a conductive phase, that is, carbon fibers into epoxy, enhances their thermal conductivity and thermal diffusivity in the transverse direction, surpassing those of pure epoxy. And the longitudinal thermal transport properties are approximately ten times greater than the transverse properties. This difference can be attributed to the efficient phonon transport along the carbon fibers in the longitudinal direction, where phonons experience minimal scattering and do not lose their energy.

In each of these materials, the thermal diffusivity and conductivity reduce as the temperature rises. This behavior can be attributed primarily to an increase in the polymer's free volume, resulting in higher thermal resistance. Consequently, the altered phonon transport within the polymer chains contributes significantly to the decrease in thermal conductance. The thermal conductivity of CFRP composite is reduced by 31\% in the transverse direction when the temperature increases from 25 \textdegree C to 100 \textdegree C. In contrast, the longitudinal direction shows a reduction of 17\% over the same temperature range. This observed difference in behavior can be potentially attributed to the presence of conductive fibers in the longitudinal direction, which offer an efficient thermal pathway for thermal transport. 

Fig~\ref{specific_heat_unaged} illustrates the variation in specific heat capacity across epoxy and CFRP composite samples, which represents the amount of energy required to raise the temperature of 1g of the specimen by 1 \textdegree C. As the temperature rises, the specific heat capacity exhibits a noticeable upward trend, consistent with the above mentioned rationale. This can be attributed to an increase in free volume that poses a challenge to elevating the system's temperature, ultimately requiring more energy. Furthermore, the specific heat capacity of the epoxy is higher than that of the CFRP composite, primarily due to their insulating nature. Insulating materials require more energy to increase their temperature by 1 degree Celsius. Conversely, the inclusion of conductive carbon fibers in the composite reduces its specific heat. 


\subsection{Effect of fiber morphology on transverse and longitudinal thermal conductivity - Computational observations}

\begin{figure}[h!]
\centering
\subfigure[]{
\includegraphics[width=7.cm]{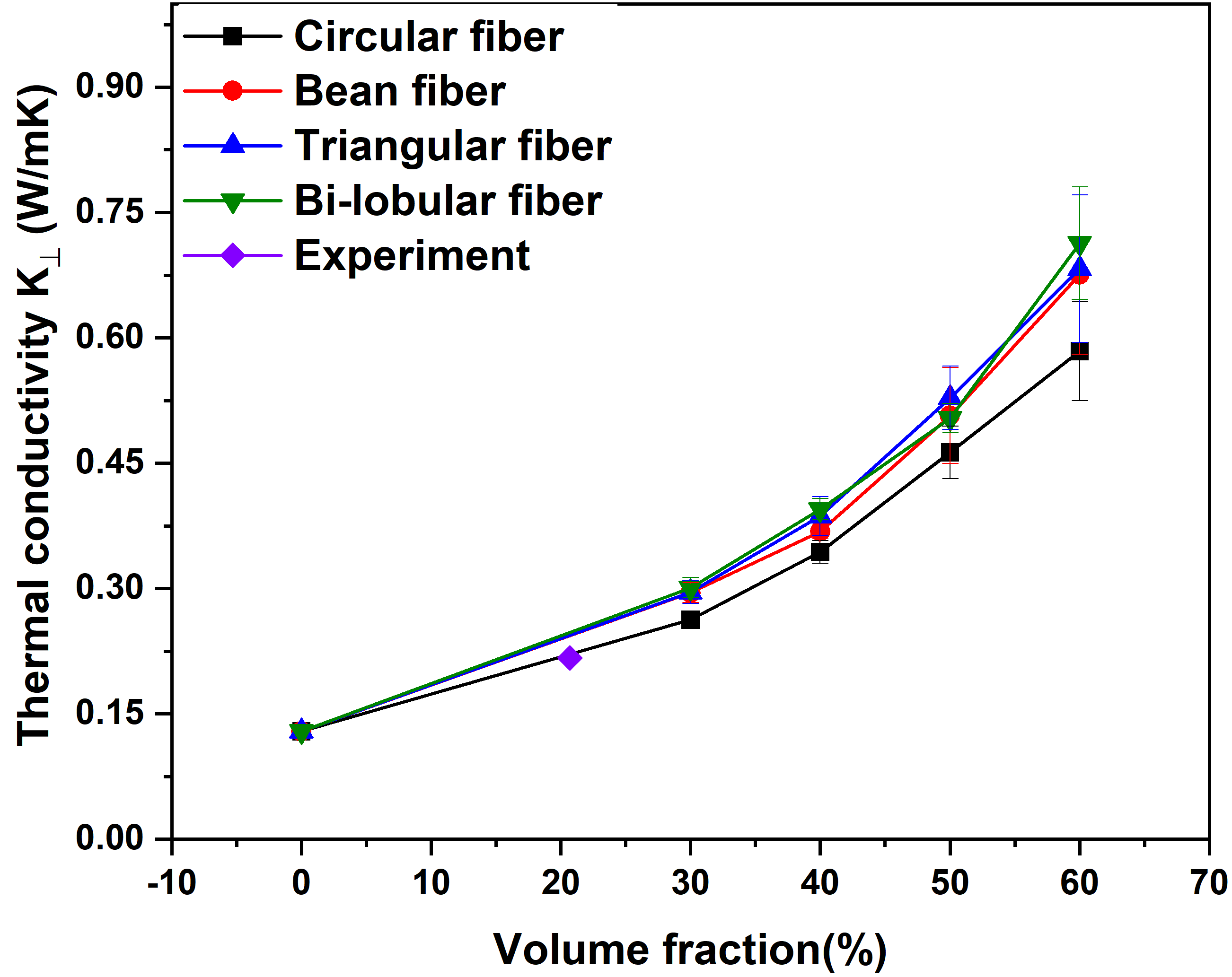}\label{transv_computation}
}
\hspace{0.2in}
\centering
\subfigure[]{
\includegraphics[width=6.75cm]{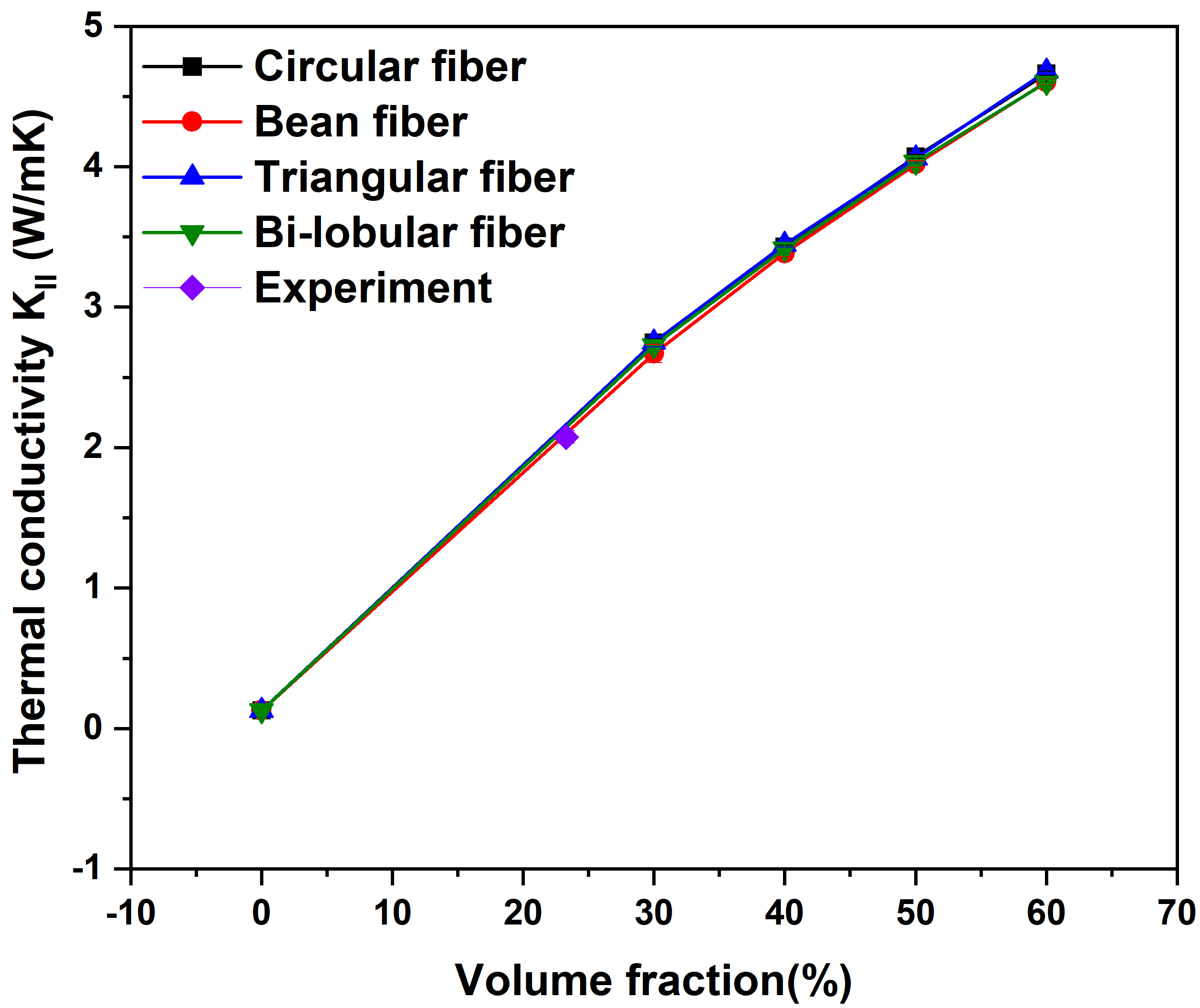}\label{long_computation} 
}
\centering
\subfigure[]{
\includegraphics[width=7.5cm]{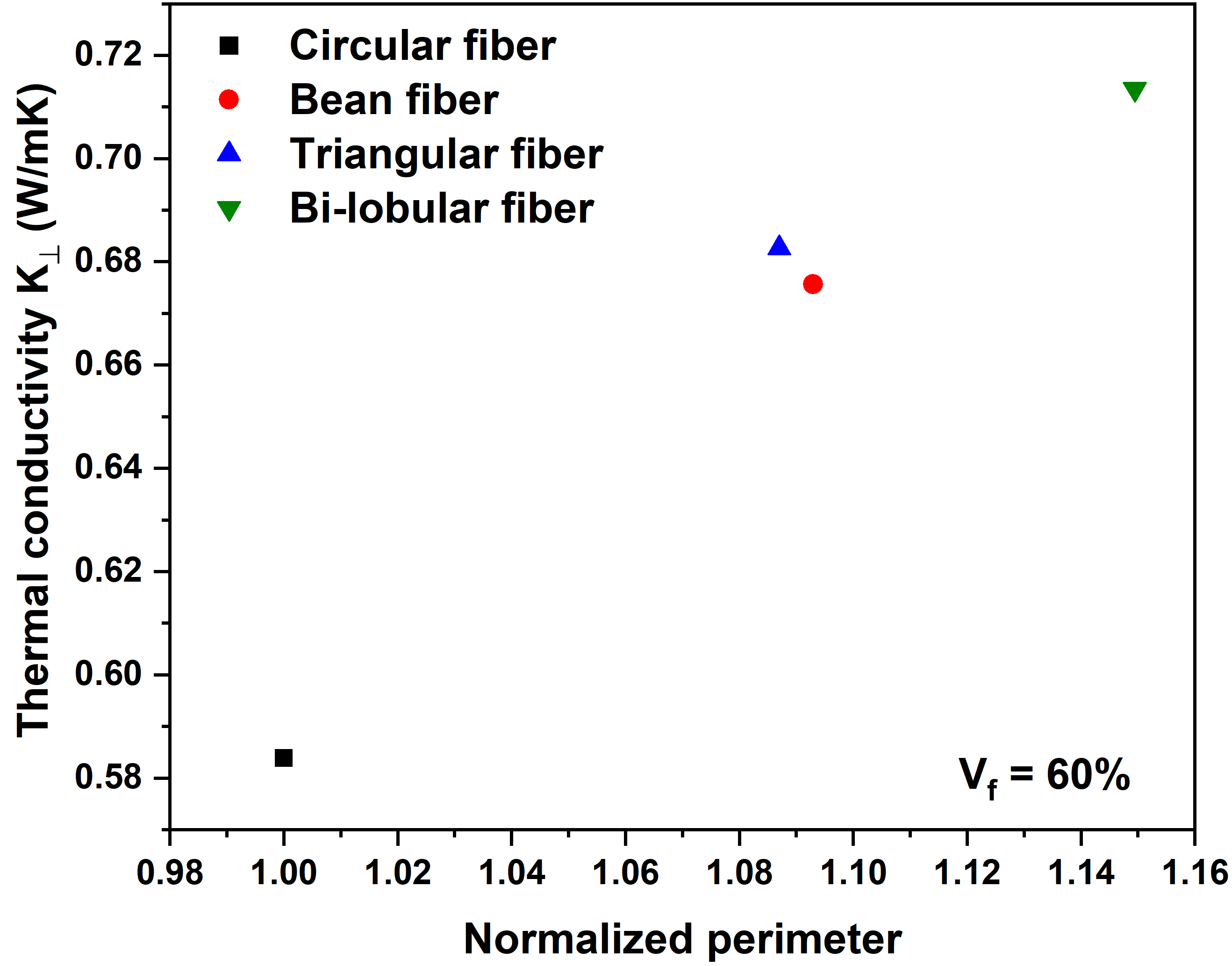}\label{conductivityPerimeter} 
}
\caption{Thermal conductivity for different fiber morphology at different volume fractions: (a) Transverse thermal conductivity (b) Longitudinal thermal conductivity. (c) Transverse thermal conductivity at 60\% fiber volume fraction for different fiber morphologies}
\end{figure}

Fig~\ref{transv_computation} and \ref{long_computation} show the transverse and longitudinal thermal conductivities of circular, bi-lobular, triangular, and bean-shaped fibers as a function of fiber volume fraction determined using computational modeling. It is observed that both transverse and longitudinal thermal conductivities increase with increasing fiber volume fraction, closely aligning with experimental results. This behavior can be attributed to the presence of larger regions of conductive phase (carbon fiber), which enhances thermal transport capabilities. Conversely, the relationship would be inverted if insulative fibers, such as glass fibers, were employed. 

The impact of fiber cross-sectional morphology on the longitudinal thermal conductivity was found to be insignificant, whereas it emerged as a critical architectural parameter for the transverse thermal conductivity. For instance, the transverse thermal conductivity of all non-circular is higher than the circular fibers because the surface area (perpendicular to the cross-section) of the non-circular fibers is higher than the circular fiber, which reduces the effective path of travel between the neighboring fibers even at lower volume fraction, which serves as an efficient phonon pathway that augments the thermal transport property. Nevertheless, the benefits derived from the increased surface area diminish when the volume fractions decrease, as the fibers become more widely spaced, resulting in phonon transport primarily occurring through the polymer chains rather than the fibers.Fig~\ref{conductivityPerimeter} shows that at fiber volume fraction of 60\%, higher fiber surface area in 3D or fiber perimeter in 2D results in increased thermal conductivity.


\section{Conclusion}\label{conc}
Several key findings have been established in this study regarding the thermal properties of epoxy and carbon fiber-reinforced epoxy composites and the impact of fiber volume fraction and architectures: 
\begin{enumerate}
    \item The thermal conductivity and thermal diffusivity of both epoxy and carbon fiber reinforced epoxy composites show a decreasing trend with increasing temperature. This can be attributed to the expansion of free volume within the material and the resulting increase in thermal resistance.
    \item The longitudinal thermal conductivity of these composites is approximately ten times greater than the transverse thermal conductivity. This stark contrast is due to the presence of a continuous heat flow path/channel offered by the fibers in the longitudinal direction, whereas the transverse direction experiences limited heat conduction due to the insulating nature of the epoxy medium. 
    \item Through numerical analysis, it was determined that an increase in the fiber volume fraction results in a corresponding increase in thermal conductivity within the composites due to the greater presence of a conductive medium. 
    \item Further, non-circular fibers were found to exhibit higher transverse thermal conductivity compared to circular fibers. This is primarily attributed to the highly connected heat pathways facilitated by the larger surface area of non-circular fibers compared to conventional fibers with the same cross-sectional areas. Conversely, the longitudinal thermal conductivity showed negligible differences between different fiber shapes. 
\end{enumerate}

These findings shed light on the thermal behavior and characteristics of epoxy and carbon fiber-reinforced epoxy composites, providing valuable insights for the optimization and design of materials with enhanced thermal properties.


\section*{Author Contributions}

{\bf Sabarinathan P Subramaniyan:} Conceptualization, Methodology, Formal Analysis, Visualization, Investigation, Writing - Original Draft. {\bf Jonathan D Boehnlein:} Investigation. {\bf Pavana Prabhakar:} Conceptualization, Methodology, Writing - Original Draft, Visualization, Verification, Supervision, Project Administration, Funding Acquisition. 


\section*{Funding}

The authors would like to acknowledge the support of the National Science Foundation (NSF) CAREER Award \# 2046476 through the Mechanics of Materials and Structures (MOMS) Program for conducting the research presented here.


\section*{Acknowledgment}
The authors express their gratitude to the Polymer Engineering Center for providing access to the Netzsch LFA 447 instrument to conduct Laser Flash Analysis. This research was partially supported by the University of Wisconsin - Madison College of Engineering Shared Research Facilities and the NSF through the Materials Science Research and Engineering Center (DMR-1720415) using instrumentation provided at the UW - Madison Materials Science Center. 
 


\bibliographystyle{unsrt}
\bibliography{references}

\clearpage

\section*{Supplementary Document}

\setcounter{section}{0}
\setcounter{figure}{0}
\setcounter{equation}{0}

\section{Volume Fraction - Thermogravimetric analysis}
\begin{figure}[h!]
\centering
\includegraphics[width=7cm]{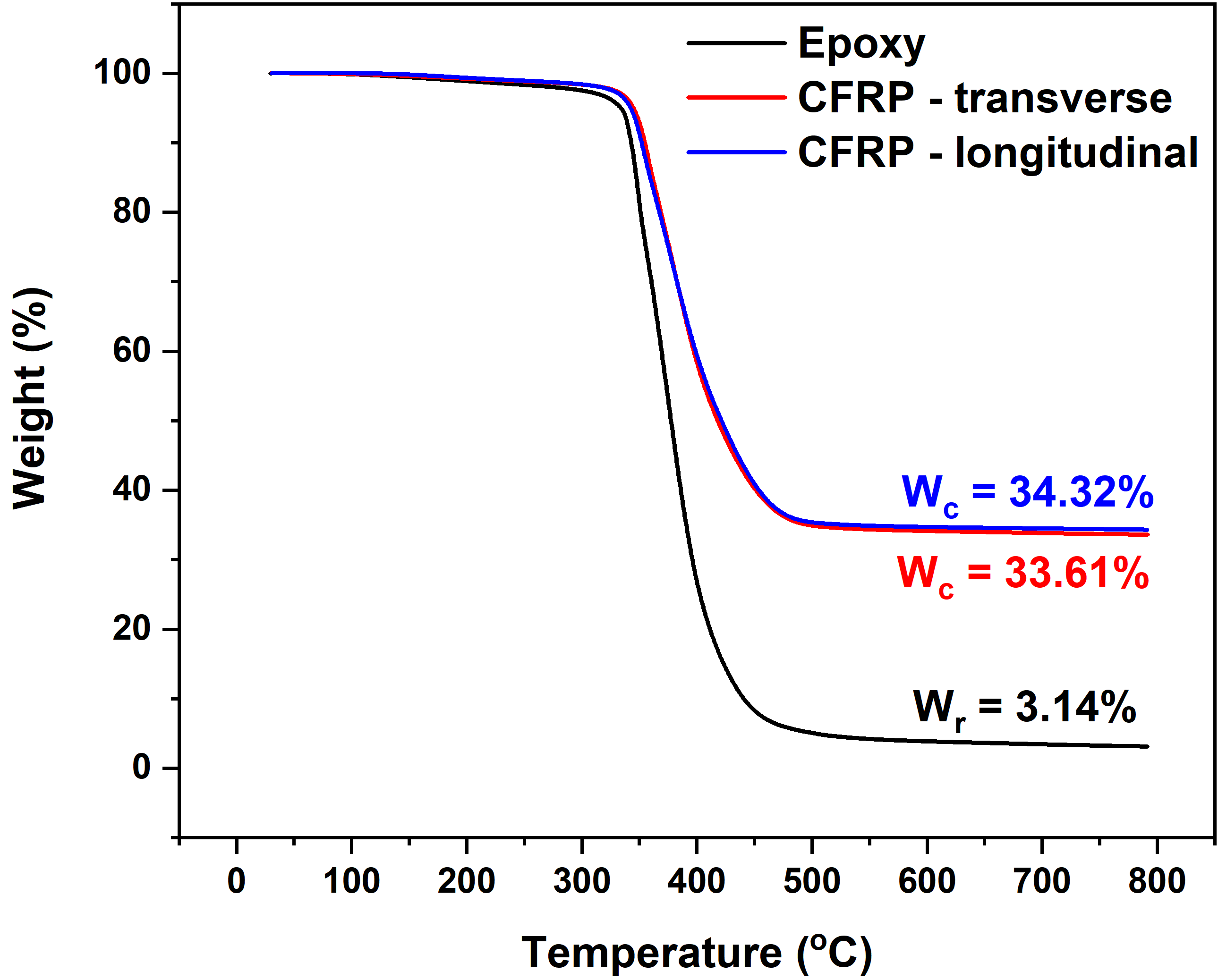}\label{fig:schemTGA}
\caption{Thermogravimetric Analysis to determine the residual weight percentage of composites and epoxy resin.}
\end{figure}

The fiber volume fraction of the fiber-reinforced polymer composites mentioned in the previous section is determined using the widely employed Thermogravimetric analysis (TGA). Determining fiber volume fraction in composite materials is critical for their characterization and performance evaluation. TGA involves heating a small composite sample, typically weighing 15 and 35 mg, in a platinum crucible under a nitrogen environment with purge flow. The sample is heated at a constant rate of 20 \textdegree C/min until it reaches a temperature of 800 \textdegree C. The TGA provides weight loss curves as shown in Figure~\ref{fig:schemTGA}, which are graphs of residual weight versus temperature. The plateau obtained after burnout at high temperatures (here 800 \textdegree C) is the weight percentage of the unburnt residue of a material. This residue is analyzed to calculate the volume fraction of carbon fiber in the composite. Figure~\ref{fig:schemTGA} shows three graphs, one for epoxy, and one each for CFRP samples fabricated to the determine transverse and longitudinal conductivity. By accounting for the contribution of individual components, this method accurately assesses fiber content in polymer composites. 

The following equations are used for calculating the fiber volume fraction $V_f$. Here, $W_c$ and $W_r$ refer to the residual weight fractions of composite and resin after burnout from TGA. $\rho_m$ and $\rho_f$ are the densities of the matrix (resin) and the fiber.

\begin{equation}\label{Wf}
    \centering
    W_f = \frac{W_c - W_r}{1 - W_r} 
\end{equation}

\begin{equation}
    \centering
    W_m = 1 - W_f
\end{equation}

\begin{equation}
    \centering
    V_f = \frac{\rho_m * W_f}{(\rho_m * W_f + \rho_f*W_m})
\end{equation}\label{Vf}

\begin{figure}[h!]
\centering
\subfigure[]{
\includegraphics[width=5.5cm]{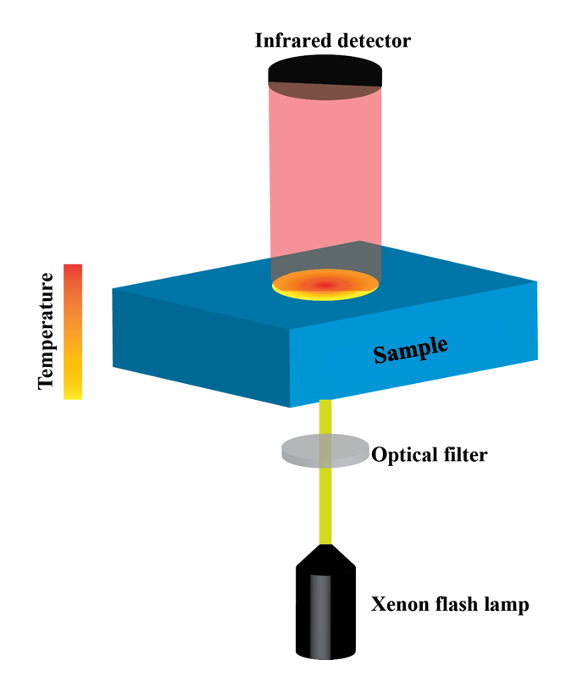}
}
\hspace{0.75in}
\centering
\subfigure[]{
\includegraphics[width=7cm]{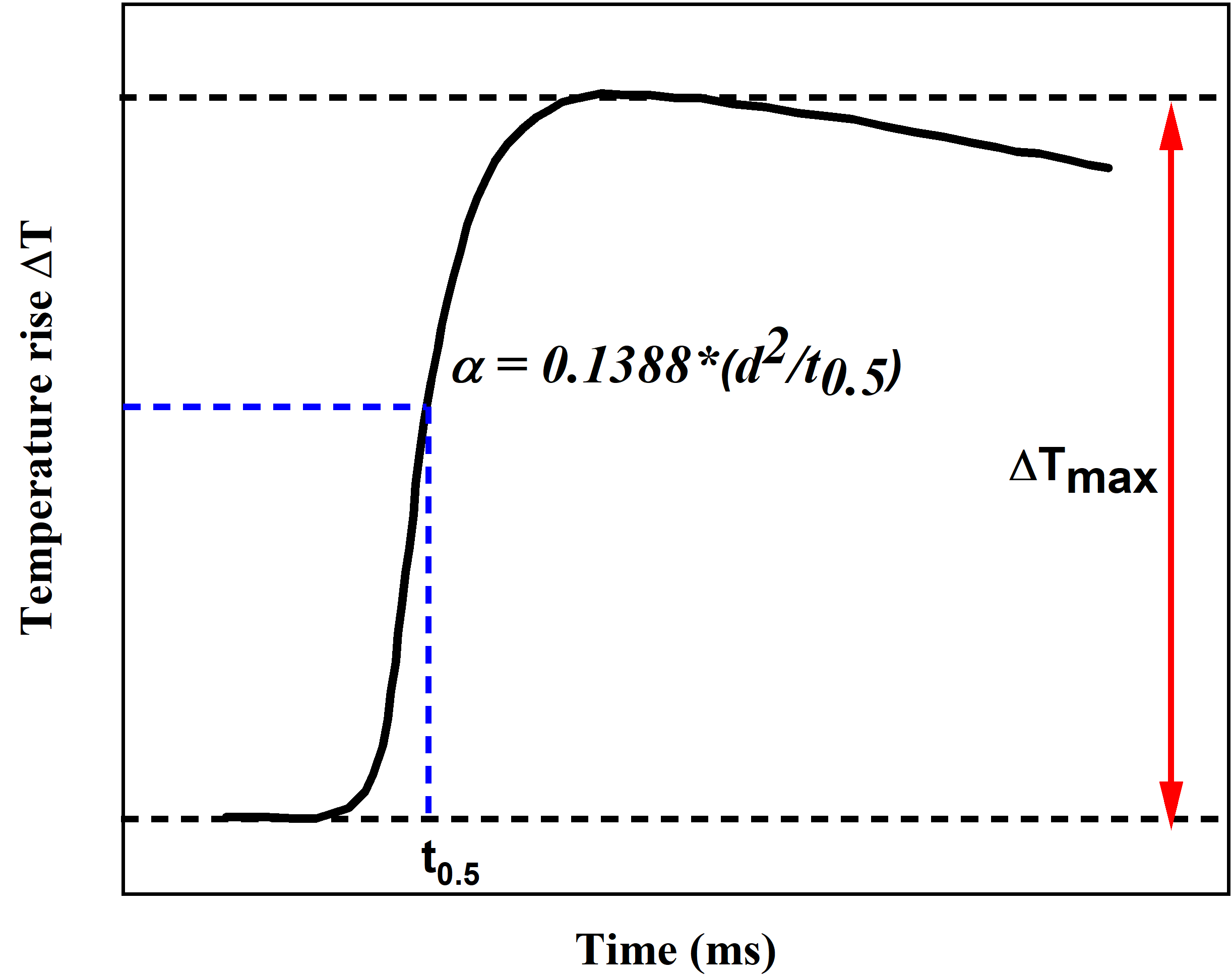} 
}
\caption{(a) Schematic illustration of laser flash analysis and (b) Thermal diffusivity measurement using Time vs Temperature rise plot}\label{LFAtest}
\end{figure}

\section{Thermal transport characterization}
Thermal properties, like thermal diffusivity, thermal conductivity, and specific heat capacity, are measured using NETZSCH LFA 447 (a Laser flash analysis technique) as schematically illustrated in Fig~\ref{LFAtest}(a) according to ASTM E1461\cite{ASTM_laser}.

The sample dimension is maintained at 10 mm x 10 mm x 1 mm and is placed in a holder inside the furnace. A light pulse of a certain width is fired on the sample’s rear face once the sample reaches the thermal equilibrium set to 30\textdegree C, and then the temperature rise is measured with respect to time as a signal using an IR detector. The thermal diffusivity $\alpha$ is calculated based on the half temperature rise, as shown in Equation~\ref{eqn:alpha}. The heat loss correction and irradiation effects are rectified using the Radiation + pulse correction method. 

Fig~\ref{LFAtest}(b) shows a schematic of the time-temperature output from the LFA tests, where,

\begin{equation}
    \alpha = 0.1388*(d^2/t_{0.5}) \label{eqn:alpha}
\end{equation}
where, $\alpha$ is the thermal diffusivity in $mm^2$/s, d is the sample thickness in mm, and $t_{0.5}$ is the half rise time.

The thermal conductivity, $\kappa$ in W/(m*K), is given by,
\begin{equation}
    \kappa = \alpha*\rho*C_p
\end{equation}
where, $\rho$ is the density of the material in g/$cm^3$ and $C_p$ is the specific heat capacity of the material in J/g/K.

Then, the material's specific heat capacity is calculated using the comparison method with a reference material with a known specific heat value. In this work, Pyroceram 9606 is used as the reference material.
The Cp of the unknown sample is measured using the following equation,
\begin{equation}
    C_p^{unknown} = \frac{T_{max}^{ref}}{T_{max}^{unknown}} * \frac{\rho^{ref}}{\rho^{unknown}}*\frac{d^{ref}}{d^{unknown}}*C_p^{ref}
\end{equation}

$T_{max}^{ref}$ and $T_{max}^{unknown}$ refer to the maximum temperature rise signal for reference samples and unknown samples, respectively, and $d_{ref}$ and $d_{unknown}$ refer to the thickness of the reference and unknown sample, respectively.

The experimentally measured fiber volume fraction of the manufactured CFRP composite and the thermal properties of epoxy and CFRP composite are used as input to the computational model described next.

\end{document}